\documentclass[aps,prb,superscriptaddress,twocolumn]{revtex4-2}
\usepackage{graphicx}
\usepackage{amsmath}
\usepackage{amssymb}
\usepackage{bm}
\usepackage{xcolor}
\usepackage{booktabs}
\definecolor{myc}{RGB}{50, 50, 141}

\usepackage[
	colorlinks=true,
	pdfstartview=Fit,
	linkcolor=blue,
	citecolor=blue,
	urlcolor=blue,
]{hyperref}

\allowdisplaybreaks[1]

\begin{document}

\title{Chiral-phonon-mediated spin current injection into an Ising superconductor}

\author{Jiayan Zhang}
\affiliation{Key Laboratory of Advanced Optoelectronic Quantum Architecture and Measurement (MOE), School of Physics, Beijing Institute of Technology, Beijing 100081, China}

\author{Gaoyang Li}
\email[]{ligy@sust.edu.cn}
\affiliation{School of Physics $\&$ Information Science, Shaanxi University of Science $\&$ Technology, Xi'an 710021, China}

\author{Yanxia Xing}
\email[]{xingyanxia@bit.edu.cn}
\affiliation{Key Laboratory of Advanced Optoelectronic Quantum Architecture and Measurement (MOE), School of Physics, Beijing Institute of Technology, Beijing 100081, China}

\date{\today}

\begin{abstract}
We theoretically propose chiral-phonon-mediated spin current injection into an Ising superconductor in a chiral-insulator/Ising-superconductor heterostructure, where nonequilibrium phonons driven by a temperature gradient transfer their angular momentum to electron spins via interfacial phonon-spin conversion.
Using the nonequilibrium Green's function formalism, we derive an expression for the injected spin current and show that it depends on the difference between the contributions from the two phonon branches with opposite circular polarizations.
We further investigate the temperature dependence of the injected spin current density at different in-plane magnetic field strengths.
In the superconducting state, the spin current density increases with temperature or in-plane magnetic field strength.
In addition to the kink at the superconducting transition, a second kink appears below the transition temperature.
The position of this kink is consistent with the onset of phonon-induced two-quasiparticle excitations as the superconducting gap decreases with increasing temperature.
We further investigate the magnetic-field dependence of the kink temperature in a temperature--magnetic-field phase diagram.
\end{abstract}

\maketitle
\section{Introduction}
The generation and control of spin currents are central objectives of spintronics~\cite{RMP2004, Bader2010}.
In electronic systems, various mechanisms enable spin-current generation and transport, including spin injection from ferromagnetic metals into normal metals~\cite{Johnson1985, Jedema2001}, the spin Hall effect~\cite{Hirsch1999, Kato2004, Valenzuela2006}, helical edge transport in quantum spin Hall systems~\cite{Bernevig2006, Konig2007}, and chirality-induced spin selectivity~\cite{Gohler2011, GAM2012}.
Near-field thermal radiation can provide another way to drive spin transport~\cite{Tang2026}.
Beyond electronic systems, magnons in magnetic insulators can also carry spin angular momentum and support spin transport~\cite{Uchida2010, Cornelissen2015, Adachi2013, LGY2022}.

Superconductors have also emerged as an important platform for spintronics~\cite{Linder2015, Bergeret2018}, where the spin angular momentum can be transported by Bogoliubov quasiparticles~\cite{Hubler2012, Wolf2013, Kato2019, Sun2023, Wakamura2014, Vargas2020, Funaki2023, Funato2022}, spin-triplet Cooper pairs~\cite{Dai2022, Bobkov2024}, or superconducting vortices~\cite{Kim2018, Meng2026}.
Previous studies of spin current injection into superconductors have primarily considered adjacent normal metals~\cite{Wolf2013}, ferromagnetic metals~\cite{Hubler2012}, or ferromagnetic insulators as spin sources~\cite{Kato2019, Sun2023, Meng2026}.
A natural question is whether alternative sources of angular momentum can induce spin current injection into a superconductor.

Circularly polarized lattice vibrations can carry angular momentum, as shown by both theoretical and experimental studies~\cite{ZLF2014, ZLF2015, ZH2025, HYZ2018, IK2023, Ueda2023, Yang2026}.
Of interest here are chiral phonons whose angular momentum is collinear with their propagation direction~\cite{Juraschek2025}.
Transport phenomena associated with chiral phonons include chirality-induced selectivity of phonon angular momenta~\cite{Ohe2024}, the phonon Edelstein effect~\cite{Takehito2025}, orbital accumulation~\cite{Yao2026, Sato2026}, and the orbital Seebeck effect~\cite{Nabei2026}.
Particularly relevant to spin current injection is the spin Seebeck effect mediated by interfacial phonon-spin conversion, in which phonon angular momentum is converted into electron spin angular momentum in an adjacent normal metal~\cite{Kim2023, Funato2024, Naoki2025, zjy2026}.
This conversion mechanism motivates exploring chiral phonons as an angular momentum source for spin current injection into superconductors.

Ising superconductors provide a platform for investigating this possibility and the influence of an in-plane magnetic field.
Ising superconductivity has been observed in two-dimensional transition-metal dichalcogenides such as MoS$_2$ and NbSe$_2$, in which strong Ising spin-orbit coupling (SOC) gives rise to valley-dependent out-of-plane spin polarization~\cite{Xi2016, Lu2015}.
This Ising SOC protects superconductivity against in-plane magnetic fields, allowing the in-plane upper critical field to substantially exceed the Pauli limit and reach approximately $52\,\text{T}$ at $1.5\,\mathrm{K}$~\cite{Saito2016}.
Within the superconducting phase, an in-plane magnetic field can modify both the quasiparticle spectrum and the spin response.
It is therefore of interest to examine how these field-induced changes affect chiral-phonon-mediated spin current injection.

\begin{figure}[t]
	\centering
	\includegraphics[scale=0.65]{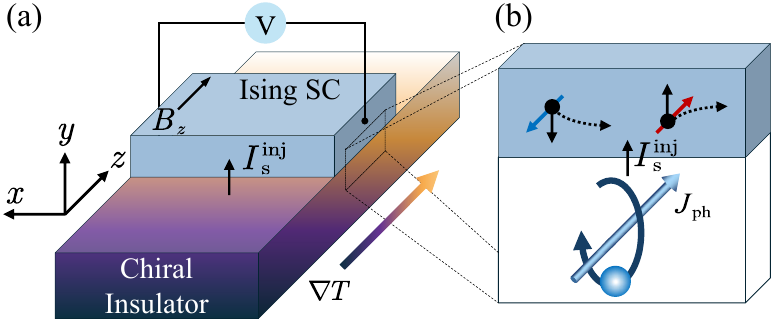}
	\caption{(a) Schematic of the proposed heterostructure.
		A temperature gradient $\nabla T$ is applied along the $z$ direction in the chiral insulator.
		$I^\text{inj}_\text{s}$ denotes the spin current injected across the interface along the $y$ direction, with its spin angular momentum polarized along the $z$ axis.
		$B_z$ is the in-plane magnetic field in the Ising superconductor.
		(b) In the chiral insulator, the angular momenta of the circularly polarized phonons are collinear with the $z$ axis. In the Ising superconductor, the trajectories [black dotted curves with arrows] illustrate how quasiparticles with opposite spins moving in opposite directions can be deflected toward the same side.
		This inverse spin Hall effect provides a means of electrically detecting the injected spin current.}
	\label{setup}
\end{figure}

In this work, we investigate spin current injection in a chiral-insulator/Ising-superconductor heterostructure, as shown by the schematic illustration in Fig.~\ref{setup}.
An Ising superconductor is placed on top of a chiral insulator.
A temperature gradient applied along the $z$ direction drives the chiral acoustic phonons in the insulator out of equilibrium.
At the interface, the circular motion of the ions in the chiral insulator interacts with electron spins in the superconductor through spin-microrotation coupling~\cite{Funato2024}.
This interaction converts phonon angular momentum into electronic spin angular momentum, thereby injecting a pure spin current into the superconductor.
In this article, we employ the nonequilibrium Green's function formalism to calculate the injected spin current to leading nonvanishing order in the interfacial coupling.
We derive an expression for the injected spin current and show that it depends on the difference between the contributions from the two phonon branches with opposite circular polarizations.
We further investigate the temperature dependence of the injected spin current density at different in-plane magnetic field strengths.

The paper is organized as follows.
Section~\ref{model} introduces the model and derives the superconducting spin susceptibility, the phonon Green's functions in the presence of a temperature gradient, and an expression for the injected spin current.
Section~\ref{result} presents and discusses the numerical results,
and Section~\ref{summary} summarizes our findings.
Appendix~\ref{Wig_tran} provides the details of the Wigner transformation of the phonon equation of motion.
Appendix~\ref{local_nm} provides a detailed derivation of the local advanced spin susceptibility in the normal state.

\section{Model and theoretical calculations}\label{model}
\subsection{Model and Hamiltonians}
The total Hamiltonian of the system is given by
\begin{equation}\label{H_tot}
	\hat{\mathcal{H}}_{\text{tot}} = 
	\hat{\mathcal{H}}_{\text{sc}} + 
	\hat{\mathcal{H}}_{\text{ph}} + 
	\hat{\mathcal{H}}_{e\text{-ph}},
\end{equation}
where $\hat{\mathcal{H}}_{\text{sc}}$ is the mean-field Hamiltonian of the Ising superconductor.
In the Nambu basis 
\begin{equation}
	\hat{\Psi}_{\bm{k}\nu} = \big[ 
	\hat{c}_{\bm{k} \nu \uparrow} , \;
	\hat{c}_{\bm{k} \nu \downarrow}  , \;
	\hat{c}^\dagger_{-\bm{k}, -\nu \uparrow} , \;
	\hat{c}^\dagger_{-\bm{k},-\nu  \downarrow} 
	\big] ^{\text{T}},
\end{equation}
$\hat{\mathcal{H}}_{\text{sc}}$ is written as
\begin{equation}
	\hat{\mathcal{H}}_{\text{sc}} = \frac{1}{2} \sum_{\bm{k},\nu} 
	\hat{\Psi}_{\bm{k}\nu}^\dagger 
	H_{\text{BdG}}(\bm{k},\nu) 
	\hat{\Psi}_{\bm{k}\nu} ,
\end{equation}
where the Bogoliubov-de Gennes (BdG) Hamiltonian near the $\bm{K}$
or $\bm{K}'$ valley, with an $s$-wave pairing potential $\Delta$, is given by~\cite{Tang2021, Li2025} 
\begin{equation}\label{H_BdG}
H_{\text{BdG}}(\bm{k},\nu) = 
\begin{bmatrix}
	h_0(\bm{k}, \nu)     &    \Delta i \sigma_y        \\ 
	- \Delta i \sigma_y  &    -h_0^\ast(-\bm{k}, -\nu)
\end{bmatrix} .
\end{equation}
In our model, the Ising superconductor lies in the $zx$ plane, with the $y$ axis normal to the interface.
Under this convention, the normal-state Hamiltonian $h_0$ is given by
\begin{equation}
	h_0(\bm{k},\nu) = \xi_{\bm{k}} \sigma_0 + \nu\beta_{\text{so}} \sigma_y - B_z \sigma_z,
\end{equation}
where $\bm{k}=(k_z, k_x)$ denotes the in-plane wavevector measured  relative to the $\bm{K}$ or $\bm{K}'$ valley, $\xi_{\bm{k}} = \hbar^2 k^2/(2m^\ast) - \mu$ is the electron dispersion relative to the chemical potential $\mu$, and $\nu=\pm 1$ is the valley index.
Here, $\sigma_0$ is the identity matrix, and $\sigma_{y,z}$ are the Pauli matrices acting in spin space.
$m^\ast$ is the effective mass.
$\beta_{\text{so}}$ denotes the Ising SOC strength.
The in-plane magnetic field gives rise to the Zeeman term $-B_z \sigma_z$, where the factor $g_\text{L} \mu_\text{B}/2$ has been absorbed into $B_z$.
Here, $g_\text{L}$ and $\mu_\text{B}$ denote the Land\'e $g$ factor and the Bohr magneton, respectively.
The superconducting gap $\Delta$ is taken to be real.
In this work, $\Delta$ is determined self-consistently following the method described in Refs.~\cite{Tang2021, Hein2026}.

The chiral phonon Hamiltonian $\hat{\mathcal{H}}_{\text{ph}}$ in  Eq.~\eqref{H_tot} is given by
\begin{equation}\label{H_ph}
	\hat{\mathcal{H}}_{\text{ph}} = \sum_{\bm{q}\lambda} \hbar\omega_{\bm{q}\lambda} 
	\big( 
	\hat{a}^\dagger_{\bm{q}\lambda} 
	\hat{a}_{\bm{q}\lambda} + 1/2 \big).
\end{equation}
Here, $\bm{q}$ denotes the phonon wavevector, $\lambda=\pm$ labels the two circular polarization branches, $\hat{a}^\dagger_{\bm{q}\lambda}(\hat{a}_{\bm{q}\lambda})$ is the phonon creation\,(annihilation) operator,
and $\hbar\omega_{\bm{q}\lambda}$ is the corresponding phonon dispersion.

The third term in Eq.~\eqref{H_tot}, $\hat{\mathcal{H}}_{e\text{-ph}}$,
accounts for the electron-phonon interaction at the interface and takes the form~\cite{Funato2024}
\begin{equation}
	\hat{\mathcal{H}}_{e\text{-ph}} = - \sum_{\bm{q}\lambda,\bm{p}} J_{\bm{pq}} 
	\big( 
	\hat{\Omega}^+_{\bm{q}\lambda} 
	\hat{s}^-_{-\bm{p}} + \hat{\Omega}^-_{-\bm{q}\bar{\lambda}} \hat{s}^+_{\bm{p}}  \big).
\end{equation}
Here, $J_{\bm{pq}}$ is the interfacial coupling strength, $\bar{\lambda}=-\lambda$, and $\hat{\Omega}^{\pm}_{\bm{q}\lambda} = \hat{\Omega}^x_{\bm{q}\lambda} \pm i \hat{\Omega}^y_{\bm{q}\lambda}$
are the ladder operators of the phonon vorticity $\hat{\bm{\Omega}}_{\bm{q}\lambda}$,
which characterizes the rotational motion of the lattice and is given by
\begin{equation}
	\hat{\bm{\Omega}}_{\bm{q}\lambda} = \sqrt{
	\frac{\hbar\omega_{\bm{q}\lambda}}
	{2 \rho_{\text{ph}} V_{\text{ph}}}} 
	(\bm{q}\times \bm{e}_{\bm{q}\lambda}) 
	\big( 
	\hat{a}_{\bm{q}\lambda} - \hat{a}^\dagger_{-\bm{q}\bar{\lambda}} \big) ,
\end{equation}
where $\rho_{\text{ph}}$ and $V_{\text{ph}}$ denote the mass density and volume of the chiral insulator, respectively,
and $\bm{e}_{\bm{q}\lambda}$ is the polarization vector of the chiral phonon.
The spin ladder operators in $\hat{\mathcal{H}}_{e\text{-ph}}$ are expressed as
\begin{subequations}\label{spin_ladder}
\begin{align}
	\hat{s}_{\bm{p}}^+ & = \hat{s}_{\bm{p}}^x + i\hat{s}_{\bm{p}}^y 
	= \hbar \sum_{\bm{k},\nu} 
	\hat{c}^\dagger_{\bm{k}\nu\uparrow}
	\hat{c}_{\bm{k}+\bm{p}\nu\downarrow} , 
	\\
	\hat{s}_{\bm{p}}^- & = \hat{s}_{\bm{p}}^x - i\hat{s}_{\bm{p}}^y 
	= \hbar \sum_{\bm{k},\nu} 
	\hat{c}^\dagger_{\bm{k}\nu\downarrow}
	\hat{c}_{\bm{k}+\bm{p}\nu\uparrow} ,
\end{align}	
\end{subequations}
where the $\alpha$-component $(\alpha=x,y,z)$ of the spin density operator is given by
\begin{equation}
	\hat{s}^\alpha_{\bm{p}} = \frac{\hbar}{2} \sum_{\bm{k},\nu} 
	\sum_{ \substack{ \sigma', \; \sigma'' \\ \in \{ \uparrow ,\downarrow \} } }  
	\hat{c}^\dagger_{\bm{k}\nu\sigma'}  [\sigma_{\alpha}]_{\sigma' \sigma''} 
	\hat{c}_{\bm{k}+\bm{p}\nu\sigma''} .
\end{equation}

\subsection{Spin current injected into the Ising superconductor}
According to the Heisenberg equation of motion, the rate of change of the $z$ component of the total spin angular momentum in the Ising superconductor is
\begin{equation}
	\frac{\partial \hat{s}^z_0}{\partial t} 
	= \frac{1}{i\hbar} 
	\big[ \hat{s}^z_0, \hat{\mathcal{H}}_{\text{tot}} \big] 
	= \hat{I}_{\text{s}}^{\text{inj}} + 
	\hat{M}_{\text{soc}}^z ,
\end{equation}
where the first term,
\begin{equation}\label{Is_inj}
	I_\text{s}^\text{inj} 
	= \frac{1}{i\hbar} \big\langle 
	\big[ \hat{s}^z_0, \hat{\mathcal{H}}_{e\text{-ph}} \big] 
	\big\rangle,
\end{equation}
represents the spin current injected into the Ising superconductor through the interfacial phonon-spin conversion,
while the second term,
\begin{equation}
	M_\text{soc}^z = \frac{1}{i\hbar} \big\langle 
	\big[ \hat{s}^z_0, \hat{\mathcal{H}}_{\text{sc}} \big] 
	\big\rangle,
\end{equation}
represents the spin torque arising from Ising SOC in the superconductor.
Since our study focuses on chiral-phonon-induced spin current injection across the chiral-insulator/Ising-superconductor interface, rather than on the effects of spin torque on spin transport within the superconductor, we consider only the spin injection contribution $I_\text{s}^\text{inj}$ in Eq.~\eqref{Is_inj}.

For convenience, we first introduce the vorticity--spin-density correlation function as
\begin{equation}
	C_{\bm{q}\lambda,\bm{p}}(\tau,\tau') 
	= -\frac{i}{\hbar}
	\big\langle \mathcal{T}_\text{c} \hat{\Omega}^+_{\bm{q}\lambda}(\tau)  \hat{s}^-_{-\bm{p}}(\tau')  \big\rangle  ,
\end{equation}
where $\tau$ and $\tau'$ are time arguments on the Keldysh contour and $\mathcal{T}_\text{c}$ is the contour-ordering operator.
In what follows, we take the interfacial coupling to be independent of wavevector, $J_{\bm{pq}} \approx J$.
Evaluating the commutator in Eq.~\eqref{Is_inj} and expressing its expectation value in terms of the vorticity--spin-density correlation function, we obtain
\begin{equation}\label{Is_inj_2}
	I^{\text{inj}}_\text{s} = 2J\hbar \sum_{\bm{q}\lambda,\bm{p}} \text{Re} 
	\big[ C^<_{\bm{q}\lambda,\bm{p}}(t,t) \big] .
\end{equation}
Here, we use the superscripts $>$, $<$, $r$ and $a$ to denote the greater, lesser, retarded, and advanced components of the correlation functions, respectively.
To obtain the injected spin current, we need to evaluate the lesser correlation function $C^<_{\bm{q}\lambda,\bm{p}}(t,t)$, which can be calculated using perturbation theory.
To this end, we introduce the contour-ordered evolution operator
$\hat{\mathcal{S}} = \mathcal{T}_\text{c} \exp \big[ -(i/\hbar) \int_\text{c} d\tau \hat{\mathcal{H}}_{e\text{-ph}}(\tau) \big] $.
We first define the phonon Green's function $D_{\bm{q}\lambda}$ and superconducting spin susceptibility $\chi_{\bm{p}}$ as
\begin{subequations}
\begin{align}
	& \hspace{0.25mm} D_{\bm{q}\lambda}(\tau,\tau') = -\frac{i}{\hbar} 
	\big\langle \mathcal{T}_\text{c}   \hat{a}_{\bm{q}\lambda}(\tau) \hat{a}^\dagger_{\bm{q}\lambda}(\tau') \big\rangle_0 , 
	\label{D_def}\\
	& \chi_{\bm{p}}(\tau,\tau') = -\frac{i}{N_{\text{S}} \hbar} 
	\big\langle \mathcal{T}_\text{c} \hat{s}^+_{\bm{p}}(\tau) \hat{s}^-_{-\bm{p}}(\tau') \big\rangle_0 , \label{chi_def}
\end{align}	
\end{subequations}
respectively, where $\langle\cdots\rangle_0$ denotes the ensemble average in the absence of interfacial coupling, and $N_\text{S}$ is the number of unit cells in the superconductor.
Expanding $C_{\bm{q}\lambda,\bm{p}}(\tau,\tau')$ to first order in $J$ and using the definitions of the phonon Green's function and superconducting spin susceptibility, we obtain
\begin{align}\label{C_contour}
	& C_{\bm{q}\lambda,\bm{p}}(\tau,\tau') = 
	-\frac{ J N_{\text{S}} \hbar \omega_{\bm{q}\lambda } L_{\bm{q}\lambda} } {2 \rho_{\text{ph}} V_{\text{ph}}} 
	\nonumber\\  
	& \quad \times
	\int_\text{c} d\tau_1 
	\big[ D_{\bm{q}\lambda}(\tau,\tau_1) + D_{-\bm{q}\bar{\lambda}}(\tau_1,\tau) \big] 
	\chi_{\bm{p}}(\tau_1,\tau') ,
\end{align}
where the polarization factor $L_{\bm{q}\lambda}$ is
\begin{equation}
	L_{\bm{q}\lambda} =  \big[ (\bm{q}\times \bm{e}_{\bm{q}\lambda})_{+}  (\bm{q}\times \bm{e}^\ast_{\bm{q}\lambda})_{-} \big] ,
\end{equation}
with $(\bm{q}\times \bm{e}_{\bm{q}\lambda})_{\pm} = (\bm{q}\times \bm{e}_{\bm{q}\lambda})_{x} \pm i(\bm{q}\times \bm{e}_{\bm{q}\lambda})_{y}$.
Applying the Langreth rules~\cite{Jishi2013, Stefanucci2025} to Eq.~\eqref{C_contour}, we obtain the real-time lesser component $C^<_{\bm{q}\lambda,\bm{p}}(t,t')$.
In the steady state, Fourier transformation with respect to the relative time $t-t'$ gives the frequency-domain correlation function $C^<_{\bm{q}\lambda,\bm{p}}(\omega)$.
Substituting this result into Eq.~\eqref{Is_inj_2}, we obtain the injected spin current as
\begin{align}\label{Is_inj_3}
	I^{\text{inj}}_\text{s} 
	& = \frac{J^2 N_{\text{S}} \hbar^2 }{\rho_{\text{ph}} V_{\text{ph}}} 
	\sum_{\bm{q}\lambda,\bm{p}} 
	\omega_{\bm{q}\lambda} L_{\bm{q}\lambda}
	\nonumber\\
	& \quad \times
	\int_{-\infty}^{\infty} \frac{d\omega}{2\pi}  \Big\{
	\text{Im} \big[ D^r_{\bm{q}\lambda}(\omega) + D^a_{-\bm{q}\bar{\lambda}}(-\omega) \big] 
	\text{Im} \big[ \chi_{\bm{p}}^<(\omega) \big]  \nonumber\\
	& \quad \quad + 
	\text{Im} \big[ D^<_{\bm{q}\lambda}(\omega) + D^>_{-\bm{q}\bar{\lambda}}(-\omega) \big] 
	\text{Im} \big[ \chi_{\bm{p}}^a(\omega) \big]
	\Big\} .
\end{align}
To evaluate Eq.~\eqref{Is_inj_3}, we need the spin susceptibility of the Ising superconductor and the nonequilibrium phonon Green's functions under a temperature gradient. These quantities are derived in the following subsections.

\subsection{Spin susceptibility of the Ising superconductor}
Substituting the spin ladder operators given in Eq.~\eqref{spin_ladder} into the definition of the spin susceptibility in Eq.~\eqref{chi_def} yields
\begin{align}\label{4-points}
	& \chi_{\bm{p}}(\tau,\tau') 
	= -\frac{i\hbar}{N_{\text{S}}} 
	\sum_{\bm{k},\bm{k}'} \sum_{\nu,\nu'}
	\nonumber\\
	& \quad \times \big\langle \mathcal{T}_\text{c}
	\hat{c}^\dagger_{\bm{k} \nu \uparrow}(\tau) 
	\hat{c}_{\bm{k}+\bm{p} \nu\downarrow}(\tau) 
	\hat{c}^\dagger_{\bm{k}' \nu'\downarrow}(\tau') 
	\hat{c}_{\bm{k}'-\bm{p} \nu'\uparrow}(\tau') 
	\big\rangle_0 .
\end{align}
To evaluate Eq.~\eqref{4-points}, we first introduce
the following normal and anomalous Green's functions
\begin{subequations}
\begin{align}
	G_{\bm{k}\nu, \sigma\sigma'}(\tau,\tau') 
	& = -\frac{i}{\hbar} \big\langle \mathcal{T}_\text{c} \hat{c}_{\bm{k}\nu\sigma}(\tau) 
	\hat{c}^\dagger_{\bm{k}\nu\sigma'}(\tau') 
	\big\rangle_0,  \label{G_de_1} 
	\\
	F_{\bm{k}\nu,\sigma\sigma'}(\tau,\tau') 
	& = -\frac{i}{\hbar} \big\langle \mathcal{T}_\text{c} \hat{c}_{\bm{k}\nu\sigma}(\tau) \hat{c}_{-\bm{k},-\nu \sigma'}(\tau') 
	\big\rangle_0, \label{F_de_1}  
	\\
	\bar{F}_{\bm{k}\nu, \sigma\sigma'}(\tau, \tau') & = -\frac{i}{\hbar} \big\langle \mathcal{T}_\text{c} \hat{c}^\dagger_{-\bm{k},-\nu\sigma}(\tau) \hat{c}^\dagger_{\bm{k}\nu\sigma'}(\tau') \big\rangle_0 \label{F_de_2} .
\end{align}	
\end{subequations}
Applying Wick's theorem~\cite{Jishi2013, Stefanucci2025} to Eq.~\eqref{4-points} and using the Green's functions defined in Eqs.~\eqref{G_de_1}-\eqref{F_de_2}, we obtain
\begin{align}
	\chi_{\bm{p}}(\tau,\tau') 
	& = -\frac{i\hbar^3}{N_{\text{S}}} \sum_{\bm{k},\nu} 
	\big[
	G_{\bm{k}+\bm{p}\nu,\downarrow\downarrow}(\tau,\tau') G_{\bm{k}\nu,\uparrow\uparrow}(\tau',\tau) 
	\nonumber\\
	& \quad - 
	F_{\bm{k}+\bm{p}\nu,\downarrow\uparrow}(\tau,\tau') 
	\bar{F}_{\bm{k}\nu,\downarrow\uparrow}(\tau',\tau)  \big] .
\end{align}
Using the Langreth rules~\cite{Jishi2013, Stefanucci2025}
and performing a Fourier transform with respect to the relative time, we obtain the advanced spin susceptibility in the frequency domain as
\begin{align}\label{chi_a_1}
	\chi_{\bm{p}}^a(\omega) 
	& = -\frac{i\hbar^2}{N_{\text{S}}} \sum_{\bm{k},\nu} \int_{-\infty}^{\infty} 
	\frac{dE}{2\pi} 
	\big[
	G_{\bm{k}+\bm{p}\nu,\downarrow\downarrow}^a(\hbar\omega+E) G_{\bm{k}\nu,\uparrow\uparrow}^<(E) 
	\nonumber\\ 
	& \quad + 
	G_{\bm{k}+\bm{p}\nu,\downarrow\downarrow}^<(\hbar\omega+E) G_{\bm{k}\nu,\uparrow\uparrow}^r(E) 
	\nonumber\\ 
	& \quad -
	F_{\bm{k}+\bm{p}\nu,\downarrow\uparrow}^a(\hbar\omega+E) 
	\bar{F}_{\bm{k}\nu,\downarrow\uparrow}^<(E) 
	\nonumber\\ 
	& \quad - 
	F_{\bm{k}+\bm{p}\nu,\downarrow\uparrow}^<(\hbar\omega+E) 
	\bar{F}_{\bm{k}\nu,\downarrow\uparrow}^r(E)  \big] .
\end{align}
To evaluate Eq.~\eqref{chi_a_1}, we need to determine
the Green's functions of the superconductor.
The retarded Green's function matrix of the Ising
superconductor is defined as
\begin{equation}
	\mathbf{G}^r_{\bm{k}\nu}(E) = \big[ (E+ i\eta) \mathbf{I}_4 - H_\text{BdG}(\bm{k},\nu)  \big]^{-1},
\end{equation}
where $\mathbf{I}_4$ is the $4 \times 4$ identity matrix and $\eta$ is a positive infinitesimal real number.
In equilibrium, the lesser Green's function is obtained from 
\begin{equation}
	\mathbf{G}_{\bm{k}\nu}^<(E) 
	= - f(E) \big[ 
	\mathbf{G}^r_{\bm{k}\nu}(E) - 
	\mathbf{G}^a_{\bm{k}\nu}(E)  \big] ,
\end{equation}
where $f(E) = [e^{E/(k_\text{B}T)}+1]^{-1}$ is the Fermi-Dirac distribution function.
Substituting the relevant Green's function components into Eq.~\eqref{chi_a_1} and summing over the valley index $\nu$,
we obtain
\begin{equation}\label{chi_a_2}
	\chi_{\bm{p}}^a(\omega) 
	= \frac{2\hbar^2}{N_{\text{S}}} \sum_{\bm{k}} \sum_{n,m}  W^{n,m}_{\bm{k},\bm{p}}
	\frac{f(E_{\bm{k},n}) - f(E_{\bm{k}+\bm{p},m})}
	{\hbar\omega + E_{\bm{k},n} - E_{\bm{k}+\bm{p},m} - i\eta} .
\end{equation}
The factor of $2$ accounts for the two valley contributions.
The four eigenvalues of the BdG Hamiltonian in Eq.~\eqref{H_BdG} are $E_{\bm{k},n} = \pm \sqrt{a_{\bm{k}} \pm \sqrt{b_{\bm{k}}} }$, with
\begin{subequations}\label{ab}
\begin{align}
	a_{\bm{k}} & = \xi_{\bm{k}}^2+B_z^2+\beta_{\text{so}}^2+\Delta^2  ,\\
	b_{\bm{k}} & = 4 \big[ \xi_{\bm{k}}^2 (B_z^2+\beta_{\text{so}}^2) + B_z^2\Delta^2  \big] .
\end{align}	
\end{subequations}
The pairs of outer and inner signs in $E_{\bm{k},n}$
are $(+,+)$, $(+,-)$, $(-,+)$, and $(-,-)$
for $n=1,2,3,4$, respectively.
The factor $W^{n,m}_{\bm{k},\bm{p}}$ in Eq.~\eqref{chi_a_2} is expressed as
\begin{equation}\label{W_kp}
	W^{n,m}_{\bm{k},\bm{p}} = 
	r^{(11)}_{\bm{k},n}  r^{(22)}_{\bm{k}+\bm{p},m} - r^{(41)}_{\bm{k},n}  r^{(23)}_{\bm{k}+\bm{p},m} ,
\end{equation}
where the coefficients $r^{(ij)}_{\bm{k},n}$ are 
\begin{equation}
	r^{(ij)}_{\bm{k},n} = (-1)^{n+1}
	\frac{N_{\bm{k}}^{(ij)}(E_{\bm{k},n})}
	{4E_{\bm{k},n}\sqrt{b_{\bm{k}}}}
\end{equation}
with the relevant numerator functions $N_{\bm{k}}^{(ij)}(E)$ given by
\begin{subequations}
\begin{align}
	N_{\bm{k}}^{(11)}(E) & = (E + \xi_{\bm{k}} - B_z)
	(E^2 - a_{\bm{k}} - 2\xi_{\bm{k}}B_z) 
	\nonumber\\ 
	& \quad + 2(\beta_{\text{so}}^2 \xi_{\bm{k}} - B_z\Delta^2) ,
	\\
	N_{\bm{k}}^{(22)}(E) & = (E + \xi_{\bm{k}} + B_z)
	(E^2 - a_{\bm{k}} + 2\xi_{\bm{k}}B_z) 
	\nonumber\\
	& \quad + 2(\beta_{\text{so}}^2 \xi_{\bm{k}} + B_z\Delta^2),
	\\
	N_{\bm{k}}^{(23)}(E) & = -\Delta (E^2 + 2B_zE + 2B_z^2 - a_{\bm{k}}) , 
	\\
	N_{\bm{k}}^{(41)}(E) & = \Delta (E^2 - 2B_zE + 2B_z^2 - a_{\bm{k}}) .
\end{align}	
\end{subequations}
Using the Sokhotski-Plemelj formula, we obtain
\begin{align}
	\text{Im} \chi_{\bm{p}}^a(\omega) 
	& = \frac{2\pi\hbar^2}{N_{\text{S}}} \sum_{\bm{k}} \sum_{n,m}  W^{n,m}_{\bm{k},\bm{p}}
	\big[ f(E_{\bm{k},n}) - f(E_{\bm{k}+\bm{p},m}) \big]
	\nonumber\\
	& \quad \times 
	\delta(\hbar\omega + E_{\bm{k},n} - E_{\bm{k}+\bm{p},m}).
\end{align}
In equilibrium, the lesser spin susceptibility is related
to the imaginary part of the advanced component by
\begin{equation}
	\chi_{\bm{p}}^<(\omega) = -2i n(\omega) \text{Im}  \chi_{\bm{p}}^a(\omega) ,
\end{equation}
where $n(\omega) = [e^{\hbar\omega/(k_\text{B}T)}-1]^{-1}$ is the Bose-Einstein distribution function.

In the expression for the injected spin current, Eq.~\eqref{Is_inj_3}, only the spin susceptibility
depends on the wavevector $\bm{p}$.
We therefore define the local advanced spin susceptibility as
\begin{equation}\label{loc_def}
	\chi^a_{\text{loc}}(\omega) = \frac{1}{N_{\text{S}}} \sum_{\bm{p}} 
	\chi^a_{\bm{p}}(\omega).
\end{equation}
To evaluate $\chi^a_{\text{loc}}$, we rewrite
the double wavevector sum over $\bm{k}$ and $\bm{p}$
as independent sums over $\bm{k}$ and
$\bm{k}'=\bm{k}+\bm{p}$.
For the two-dimensional parabolic dispersion,
each wavevector sum is converted into an energy integral using
\begin{equation}
	\frac{1}{N_{\text{S}}} \sum_{\bm{k}} \, (\cdots) 
	\ \  \longrightarrow \ \ 
	\zeta(\mu) A_{\text{cell}}
	\int_{-\mu}^{\infty} d \xi \, (\cdots)  ,
\end{equation}
where $A_{\text{cell}}$ denotes the unit-cell area of the Ising superconductor 
and $\zeta(\mu) = m^\ast/(2\pi\hbar^2)$ is the constant density of states per unit area for a two-dimensional parabolic dispersion.
Applying this conversion yields
\begin{align}
	&  \text{Im} \chi^a_{\text{loc}}(\omega) 
	= 2\pi\hbar^2 \zeta^2 A_{\text{cell}}^2  \sum_{n,m}
	\int_{-\mu}^{\infty} d\xi \int_{-\mu}^{\infty} d\xi'
	\; W^{n,m}_{\xi,\xi'}
	\nonumber\\   
	&  \quad \times
	\big[ f(E_{\xi,n}) - f(E_{\xi',m}) \big]
	\delta(\hbar\omega + E_{\xi,n} - E_{\xi',m}) .
\end{align}

\subsection{Phonon Green's functions under a temperature gradient}
Electromagnetic fields and temperature gradients can modify the transport properties of a system.
Electromagnetic fields couple directly to charge carriers. In a tight-binding description, the orbital coupling to a vector potential can be introduced through the Peierls substitution~\cite{Peierls1933},
and the response can be calculated from the resulting perturbation Hamiltonian within linear response theory~\cite{Kubo1956}.
In contrast, a temperature gradient acts as a statistical force that modifies the carrier distribution, rather than as an explicit mechanical perturbation of the microscopic Hamiltonian.
Therefore, its response requires a suitable formulation within linear response theory, and a naive application of the theory leads to unphysical divergences near zero temperature~\cite{Agarwalla2011}.
Luttinger proposed introducing an auxiliary gravitational field to study the transport properties of a system subjected to a temperature gradient~\cite{Luttinger1964}.
In this approach, a perturbation Hamiltonian is constructed by coupling the auxiliary gravitational field to the local energy density.
The response to this auxiliary field is first calculated, and the response to a temperature gradient is then obtained by replacing the auxiliary-field gradient with the corresponding temperature gradient~\cite{Luttinger1964, Tatara2015, Tatara2015_2, Saleem2024}.
This auxiliary-field approach has been widely used to investigate thermal transport in electronic~\cite{Michaeli2009, Michaeli2009_2, Eich2014, Qin2011, Kapustin2021, Schwiete2014}, magnonic~\cite{Matsumoto2011, Mook2014, Neumann2022}, and phononic systems~\cite{Qin2012}.
We now use this auxiliary-field approach to calculate the phonon Green's functions in the presence of a temperature gradient.

To introduce the auxiliary field, we first rewrite the chiral phonon Hamiltonian in Eq.~\eqref{H_ph} in terms of the local phonon energy density.
We use the following Fourier transformations for the phonon creation and annihilation operators
\begin{subequations}
\begin{align}
	\hat{a}_{\bm{q}\lambda}^\dagger & = \frac{1}{\sqrt{N_{\text{ph}}}} \sum_i e^{i\bm{q}\cdot\bm{r}_i} \hat{a}^\dagger_{i\lambda} ,\\
	\hat{a}_{\bm{q}\lambda} & = \frac{1}{\sqrt{N_{\text{ph}}}} \sum_i e^{-i\bm{q}\cdot\bm{r}_i} \hat{a}_{i\lambda} ,
\end{align}	
\end{subequations}
where $N_\text{ph}$ denotes the number of unit cells in the chiral insulator.
Omitting the zero-point contribution, we can rewrite the phonon Hamiltonian as
\begin{equation}
	\hat{\mathcal{H}}_{\text{ph}} = \sum_i \hat{\epsilon}_i,
\end{equation}
where $\hat{\epsilon}_i$ is the local phonon energy operator assigned to lattice site $i$, defined symmetrically as
\begin{equation}
	\hat{\epsilon}_i = \frac{1}{2} \sum_{j,\lambda} \Big(
	h^\lambda_{ij} \hat{a}^\dagger_{i\lambda} \hat{a}_{j\lambda} + 
	h^\lambda_{ji} \hat{a}^\dagger_{j\lambda} \hat{a}_{i\lambda}
	\Big)
\end{equation}
with
\begin{equation}
	h^\lambda_{ij} = \frac{1}{N_\text{ph}} \sum_{\bm{q}} \hbar\omega_{\bm{q}\lambda} 
	e^{i\bm{q}\cdot(\bm{r}_i-\bm{r}_j)} .
\end{equation}
The corresponding phonon energy density is
\begin{align}
	\hat{h}_{\text{ph}}(\bm{r}) & = \sum_i \delta(\bm{r}-\bm{r}_i)
	\hat{\epsilon}_i  \nonumber\\
	& = \frac{1}{2} \sum_{ij,\lambda} \big[
	\delta(\bm{r}-\bm{r}_i) + \delta(\bm{r}-\bm{r}_j)  \big]
	h^\lambda_{ij} \hat{a}^\dagger_{i\lambda} \hat{a}_{j\lambda} .
\end{align}
In this article, the auxiliary gravitational field is denoted by $\phi(\bm{r})$.
Coupling this field to the local phonon energy density gives the field-dependent Hamiltonian
\begin{align}
	\hat{\mathcal{H}}^{\phi}_{\text{ph}} = \int d^3 \bm{r}
	\phi(\bm{r}) \hat{h}_{\text{ph}}(\bm{r})   
	= \frac{1}{2} \sum_{ij,\lambda} 
	\big( \phi_i + \phi_j \big)
	h^\lambda_{ij} \hat{a}^\dagger_{i\lambda} \hat{a}_{j\lambda} ,
\end{align}
where $\phi_i\equiv\phi(\bm{r}_i)$.
The unperturbed Hamiltonian is recovered by setting $\phi(\bm{r})=1$.
The phonon annihilation operator obeys the Heisenberg equation of motion
\begin{align}
	i\hbar \partial_t \hat{a}_{i\lambda} 
	= \big[ \hat{a}_{i\lambda}, \hat{\mathcal{H}}^\phi_{\text{ph}} \big] 
	= \frac{1}{2} \sum_n \big( \phi_i + \phi_n \big) h^\lambda_{in} \hat{a}_{n\lambda}.
\end{align}
We adopt the following form of the contour-ordered phonon Green's function in the lattice representation~\cite{Peano2015}
\begin{equation}
	D_{ij,\lambda}(\tau,\tau') = -\frac{i}{\hbar} 
	\big\langle \mathcal{T}_\text{c} \hat{a}_{i\lambda}(\tau) \hat{a}^\dagger_{j\lambda}(\tau') \big\rangle_0  ,
\end{equation}
which is the Fourier-transformed form of $D_{\bm{q}\lambda}(\tau,\tau')$ defined in Eq.~\eqref{D_def}.
Differentiating $D_{ij,\lambda}(\tau,\tau')$ with respect to the contour time $\tau$ and including a phonon self-energy to describe scattering or relaxation, we obtain the equation of motion
\begin{align}
	& i\hbar \partial_{\tau} D_{ij,\lambda}(\tau,\tau') 
	- \sum_n \tilde{h}^\lambda_{in} D_{nj,\lambda}(\tau,\tau')  
	= \delta_{ij} \delta(\tau,\tau') 
	\nonumber\\
	& \quad + 
	\sum_n \int_\text{c} d\tau_1 \Sigma_{in,\lambda}(\tau,\tau_1) D_{nj,\lambda}(\tau_1,\tau'),
\end{align}
where
\begin{equation}
	\tilde{h}^\lambda_{ij} = \frac{1}{2} (\phi_i + \phi_j) h^\lambda_{ij}.
\end{equation}
Here, $\Sigma_{in,\lambda}$ is the phonon self-energy, whose form is left unspecified at this stage.
In the steady state, we apply the Langreth rules~\cite{Jishi2013, Stefanucci2025} and then Fourier transform with respect to the relative time $t-t'$ to obtain
\begin{subequations}\label{D_r<}
\begin{align}
	& \hbar\omega D^{r/a}_{ij,\lambda}(\omega) 
	- \sum_n \tilde{h}^\lambda_{in} D^{r/a}_{nj,\lambda}(\omega) 
	\nonumber\\
	& \hspace{2.7cm} 
	= \delta_{ij} + \sum_n 
	\Sigma^{r/a}_{in,\lambda}(\omega) D^{r/a}_{nj,\lambda}(\omega)   , 
	\\
	& \hbar\omega D^<_{ij,\lambda}(\omega) 
	- \sum_n \tilde{h}^\lambda_{in} D^<_{nj,\lambda}(\omega)  
	\nonumber\\
	& \ \ = 
	\sum_n \big[ 
	\Sigma^r_{in,\lambda}(\omega) D^<_{nj,\lambda}(\omega) 
	+ \Sigma^<_{in,\lambda}(\omega) D^a_{nj,\lambda}(\omega) \big]  .
\end{align}	
\end{subequations}
A spatially varying auxiliary field breaks translational invariance.
We therefore perform a Wigner transformation on the phonon Green's function and the equations of motion.
For two lattice positions $\bm{r}_i$ and $\bm{r}_j$, we introduce the center coordinate $\bm{R} = (\bm{r}_i + \bm{r}_j)/2$ and the relative coordinate $\bm{d} = \bm{r}_i - \bm{r}_j$.
We write $D_{ij,\lambda}(\omega)$ as $D_{\lambda}(\bm{R},\bm{d},\omega)$ and define its spatial Wigner transform by
\begin{equation}
	D_{\bm{q}\lambda}(\bm{R}, \omega) 
	= \sum_{\bm{d}} e^{-i\bm{q}\cdot\bm{d}}
	D_{\lambda}(\bm{R},\bm{d},\omega) .
\end{equation}
We assume that the auxiliary field varies slowly on microscopic length scales and retain terms only to first order in spatial gradients.
Applying the Wigner transformation to Eq.~\eqref{D_r<}, we obtain the following Dyson and Keldysh equations [see Appendix~\ref{Wig_tran} for details]
\begin{subequations}
\begin{align}
	& \hspace{-1.5mm} \big[ \hbar\omega  - \phi \hbar\omega_{\bm{q}\lambda} -
	\Sigma^{r/a}_{\bm{q}\lambda}(\bm{R},\omega) \big] 
	\star D^{r/a}_{\bm{q}\lambda}(\bm{R},\omega) 
	= 1 , \label{D_r} 
	\\
	& \hspace{-1.5mm} D^{<}_{\bm{q}\lambda}(\bm{R}, \omega) = 
	D^r_{\bm{q}\lambda}(\bm{R},\omega) \star
	\Sigma^{<}_{\bm{q}\lambda}(\bm{R}, \omega) \star 
	D^{a}_{\bm{q}\lambda}(\bm{R},  \omega)  \label{D_<} .
\end{align}
\end{subequations}
Here, $\phi\equiv\phi(\bm{R})$. To first order in spatial gradients, the star product $\star$ is defined as
\begin{equation}
	A \star B = A B + \frac{i}{2} 
	\big( \nabla_{\bm{R}} A \cdot \nabla_{\bm{q}} B -
	\nabla_{\bm{q}} A \cdot \nabla_{\bm{R}} B \big) .
\end{equation}
In the present work, we model phonon relaxation phenomenologically using a phonon lifetime $\tau_{\text{ph}}$, so that the spectral half-width is given by
\begin{equation}
	\Gamma_{\text{ph}} = \hbar/(2\tau_{\text{ph}}) .
\end{equation}
We approximate the retarded and advanced phonon self-energies as
\begin{equation}\label{self_ph}
	\Sigma^r_{\lambda} = (\Sigma^a_{\lambda})^\dagger = -i\Gamma_{\text{ph}} .
\end{equation}
Substituting these self-energies into Eq.~\eqref{D_r} yields, to first order in the gradient expansion,
\begin{equation}
	D^{r/a}_{\bm{q}\lambda}(\bm{R},\omega) = 
	\big[ \hbar\omega - \phi(\bm{R}) \hbar\omega_{\bm{q}\lambda} \pm i \Gamma_{\text{ph}} \big]^{-1}.
\end{equation}
Here, the upper (lower) sign corresponds to the retarded (advanced) component.
The corresponding phonon spectral function is
\begin{align}
	A_{\bm{q}\lambda}(\bm{R},\omega) 
	& = i \big[ D^{r}_{\bm{q}\lambda}(\bm{R},\omega) 
	- D^{a}_{\bm{q}\lambda}(\bm{R},\omega) \big]  \nonumber\\
	& = \frac{2\Gamma_{\text{ph}}}
	{\big[ \hbar\omega - \phi(\bm{R}) \hbar\omega_{\bm{q}\lambda} \big]^2 + \Gamma_{\text{ph}}^2} .
\end{align}
For the lesser phonon self-energy, we adopt the ansatz
\begin{equation}\label{self_ph_<}
	\Sigma_{\lambda}^<(\bm{R},\omega) = - 2 i \tilde{n}(\bm{R}, \omega) \Gamma_{\text{ph}} ,
\end{equation}
where $\tilde{n}(\bm{R},\omega)$ is the auxiliary-field-modified Bose-Einstein distribution function given by~\cite{Michaeli2009, Eich2014}
\begin{equation}
	\tilde{n}(\bm{R},\omega) 
	= n(\bm{R},\omega/\phi)
	= \big[e^{\hbar\omega / ( \phi k_{\text{B}} T )}-1 \big]^{-1} .
\end{equation}
Substituting Eq.~\eqref{self_ph_<} into the Keldysh equation in Eq.~\eqref{D_<}, we obtain
\begin{align}
	& D^<_{\bm{q}\lambda}(\bm{R},\omega) 
	= -i A_{\bm{q}\lambda} \tilde{n}
	\nonumber\\ & \quad
	- \frac{i\hbar}{2} A^2_{\bm{q}\lambda}   
	\big(
	\omega_{\bm{q}\lambda} \nabla_{\bm{R}} \phi \cdot \nabla_{\bm{q}} \tilde{n}  - 
	\phi \nabla_{\bm{q}} \omega_{\bm{q}\lambda} \cdot \nabla_{\bm{R}} \tilde{n} \big) ,
\end{align}
where $\nabla_{\bm{q}} \tilde{n}(\bm{R},\omega) = 0$ and
\begin{align}
	\nabla_{\bm{R}} \tilde{n}(\bm{R},\omega)
	& = \frac{e^{\hbar\omega / ( \phi k_{\text{B}} T )}}
	{\big[e^{\hbar\omega / ( \phi k_{\text{B}} T )} - 1\big]^2}
	\frac{\hbar\omega}{k_{\text{B}}T \phi^2} \nabla_{\bm{R}} \phi
	\nonumber\\
	& = - \frac{\omega}{\phi} (\nabla_{\bm{R}} \phi)
	\frac{\partial}{\partial \omega}
	\tilde{n}(\bm{R},\omega) .
\end{align}	
To obtain the thermal response, we replace the auxiliary-field gradient by the temperature gradient, $\nabla_{\bm{R}}\phi\rightarrow\nabla T/T$, and subsequently set $\phi=1$~\cite{Michaeli2009}.
In our calculations, we assume that the temperature of the Ising superconductor is spatially uniform and equal to the local reference temperature of the chiral insulator at the interface.
We therefore evaluate the phonon Green's function using this reference temperature $T$, obtaining
\begin{align}
	D^<_{\bm{q}\lambda}(\omega) 
	& = -i n(\omega) A_{\bm{q}\lambda}(\omega) 
	\nonumber\\ 
	& \quad - \frac{i\hbar\omega}{2T} 
	A_{\bm{q}\lambda}^2(\omega)  
	\big( \nabla_{\bm{q}} \omega_{\bm{q}\lambda} \cdot \nabla T \big)
	\frac{\partial n(\omega)}{\partial \omega}.
\end{align}
Using the identity $A_{\bm{q}\lambda}(\omega) = i \big[ D^>_{\bm{q}\lambda}(\omega) - D^<_{\bm{q}\lambda}(\omega) \big]$, we obtain the greater component
\begin{align}
	D^>_{\bm{q}\lambda}(\omega) & = -i [ n(\omega) + 1] A_{\bm{q}\lambda}(\omega) 
	\nonumber\\
	& \quad - \frac{i\hbar\omega}{2T} A_{\bm{q}\lambda}^2(\omega)  
	\big( \nabla_{\bm{q}} \omega_{\bm{q}\lambda} \cdot \nabla T \big)
	\frac{\partial n(\omega)}{\partial \omega} .
\end{align}

\subsection{Expression for the injected spin current}
Substituting the expressions derived above for the superconducting spin susceptibility and the phonon Green's functions into Eq.~\eqref{Is_inj_3}, we obtain
\begin{align}\label{Is_inj_4}
	I^{\text{inj}}_\text{s} 
	& = - \frac{J^2 N_{\text{S}}^2 \hbar^3 }
	{2\rho_\text{ph} V_\text{ph} T} \sum_{\bm{q}\lambda}  \omega_{\bm{q}\lambda}  L_{\bm{q}\lambda}
	\big( \nabla_{\bm{q}} \omega_{\bm{q}\lambda} \cdot \nabla T \big)
	\nonumber\\ 
	& \quad \times
	\int_{-\infty}^{\infty} \frac{d\omega}{2\pi} \omega
	\big[A_{\bm{q}\lambda}^2(\omega) + A_{\bm{q}\lambda}^2(-\omega)\big]
	\text{Im} \chi^a_\text{loc}(\omega) 
	\frac{\partial n(\omega)}{\partial \omega} .
\end{align}
For simplicity, we choose the circular polarization vectors of the chiral phonons as
$\bm{e}_{\bm{q}\lambda} = \bm{e}_{\pm} = [1, \pm i, 0]^\text{T}/\sqrt{2}$~\cite{Funato2024},
from which the polarization factors $L_{\bm{q}\lambda}$ can be evaluated.
Since the temperature gradient is applied along the $z$ direction, $\nabla_{\bm{q}} \omega_{\bm{q}\lambda} \cdot \nabla T = v_{\bm{q}\lambda}^z \big( \partial T / \partial z \big)$, where $v_{\bm{q}\lambda}^z
= \partial\omega_{\bm{q}\lambda} / \partial q_z$
is the $z$ component of the phonon group velocity.
The sum in Eq.~\eqref{Is_inj_4} is unchanged by the relabeling $(\bm{q},\lambda)\rightarrow(-\bm{q},\bar{\lambda})$.
We can therefore average the original and relabeled expressions.
In the term containing $A_{\bm{q}\lambda}^2(-\omega)$, we then change the integration variable from $\omega$ to $-\omega$ and use the fact that the frequency derivative of the Bose distribution is even.
With these simplifications, Eq.~\eqref{Is_inj_4} reduces to
\begin{align}\label{Is_inj_5}
	I^{\text{inj}}_\text{s} 
	& = \frac{J^2 N_{\text{S}}^2 \hbar^3 }{2\rho_\text{ph} V_\text{ph} T}
	\frac{\partial T}{\partial z} 
	\sum_{\bm{q}\lambda}  \lambda q_z^2  \omega_{\bm{q}\lambda} 
	v_{\bm{q}\lambda}^z 
	\nonumber\\ 
	& \quad \times
	\int_{-\infty}^{\infty} \frac{d\omega}{2\pi} \omega 
	\text{Im} \big[
	\chi^a_\text{loc}(\omega) - \chi^a_\text{loc}(-\omega) 
	\big]
	A_{\bm{q}\lambda}^2(\omega)
	\frac{\partial n(\omega)}{\partial \omega} .
\end{align}
The polarization sum in Eq.~\eqref{Is_inj_5} can be written explicitly as the difference between the contributions from the two circular polarizations
\begin{align}\label{q_sum}
	& \sum_{\bm{q}\lambda}  \lambda q_z^2  \omega_{\bm{q}\lambda} 
	v_{\bm{q}\lambda}^z  A_{\bm{q}\lambda}^2(\omega) 
	\nonumber\\ 
	& = \sum_{\bm{q}} q_z^2  
	\big[ \omega_{\bm{q}+} v_{\bm{q}+}^z  A_{\bm{q}+}^2(\omega) 
	- \omega_{\bm{q}-} v_{\bm{q}-}^z  A_{\bm{q}-}^2(\omega) \big] .
\end{align}
In this work, we adopt the chiral phonon dispersion reported in Ref.~\cite{Kishine2020}.
For phonons propagating along the chiral axis (the $z$ axis in our setup), the two circular polarization branches are generally nondegenerate, $\omega_{\bm{q}+} \neq \omega_{\bm{q}-}$, allowing for a nonvanishing contribution to the wavevector summation in Eq.~\eqref{q_sum}.
Moreover, the phonon dispersion used here lacks inversion symmetry within each polarization branch, $\omega_{\bm{q}\lambda}\neq\omega_{-\bm{q}\lambda}$, while preserving time-reversal symmetry, $\omega_{\bm{q}\lambda} = \omega_{-\bm{q}\bar{\lambda}}$.
Therefore, the phonon group velocities and spectral functions satisfy
\begin{equation}
	v^z_{\bm{q}\lambda} = -v^z_{-\bm{q}\bar{\lambda}} ,
	\qquad
	A^2_{\bm{q}\lambda}(\omega) = A^2_{-\bm{q}\bar{\lambda}}(\omega) .
\end{equation}
We convert the sum over the phonon wavevector $\bm{q}$ into an integral according to
\begin{equation}
	\frac{1}{V_\text{ph}} \sum_{\bm{q}} \, (\cdots)
	\ \  \longrightarrow \ \
	\int \frac{d^3\bm{q}}{(2\pi)^3} \, (\cdots).
\end{equation}
Since the phonon dispersion chosen in this work depends only on the longitudinal wavevector component $q_z$~\cite{Kishine2020}, the transverse wavevector integrations can be carried out first.
Exploiting the above symmetry relations to restrict the remaining integration to the positive $q_z$ region, we obtain
\begin{align}\label{Is_inj_6}
	I^{\text{inj}}_\text{s} & = \frac{J^2 \hbar^3 N^2_{\text{S}}}{\rho_{\text{ph}} c_0^2 T} \frac{\partial T}{\partial z}
	\nonumber\\*
	& \quad \times
	\int_{-\infty}^{\infty} \frac{d\omega}{2\pi} \omega  
	\text{Im} \big[ 
	\chi^a_{\text{loc}}(\omega) - \chi^a_{\text{loc}}(-\omega)   \big]
	\frac{\partial n(\omega)}{\partial \omega} \nonumber
	\\*
	& \quad \times 
	\int_{0}^{\Lambda} \frac{d q_z}{2\pi}
	q_z^2
	\big[ 
	\omega_{q_z+} v_{q_z+}^z A_{q_z+}^2(\omega) \nonumber\\*
	& \hspace{3.6cm}
	- \omega_{q_z-} v_{q_z-}^z A_{q_z-}^2(\omega)
	\big] , 
\end{align}
where $\Lambda $ is the $q_z$ integration cutoff and $c_0$ is the lattice constant of the chiral insulator.

\section{Numerical Results}\label{result}
To numerically evaluate Eq.~\eqref{Is_inj_6}, we specify the relevant model parameters as follows.
For the Ising superconductor, we take the zero-field transition temperature to be $T_{\text{c}0} = 8\,\text{K}$~\cite{Lu2015}, corresponding to a zero-temperature superconducting gap of $\Delta_0 = 1.216\,\text{meV}$.
The chemical potential and Ising SOC strength are set to $\mu=200\,\text{meV}$ and $\beta_{\text{so}}=5\Delta_0$, respectively.
Taking MoS$_2$ as the Ising superconducting material, we use
an effective mass of $m^\ast = 0.7m_e$~\cite{Pisoni2018} and a unit-cell area of $A_\text{cell} = 0.0865\,\text{nm}^2$~\cite{Roldan2014}.
The contact area between the Ising superconductor and the chiral insulator is taken to be $A_\text{sc} = 10^6\,\mu\text{m}^2$, yielding a total number of unit cells of $N_\text{S} = A_\text{sc} / A_\text{cell} \simeq 10^{13}$. 

For the chiral insulator, we adopt a mass density of $\rho_\text{ph} = 5.029 \, \text{g}/\text{cm}^3$~\cite{Kishine2020}, a phonon lifetime of $\tau_\text{ph} = 1\,\text{ns}$, and a lattice constant of $c_0 = 0.1\,\text{nm}$.
The temperature gradient along the $z$ direction is set to $\partial T / \partial z = 10^5 \, \text{K}/\text{m}$.
The wavevector cutoff is chosen as $\Lambda = 3 \,\text{nm}^{-1}$.
Following Ref.~\cite{Funato2024}, the dimensionless interfacial electron-phonon coupling strength is taken to be $J^2 \simeq 1/N_\text{S}$.

\begin{figure}[t]
	\centering
	\includegraphics[scale=0.61]{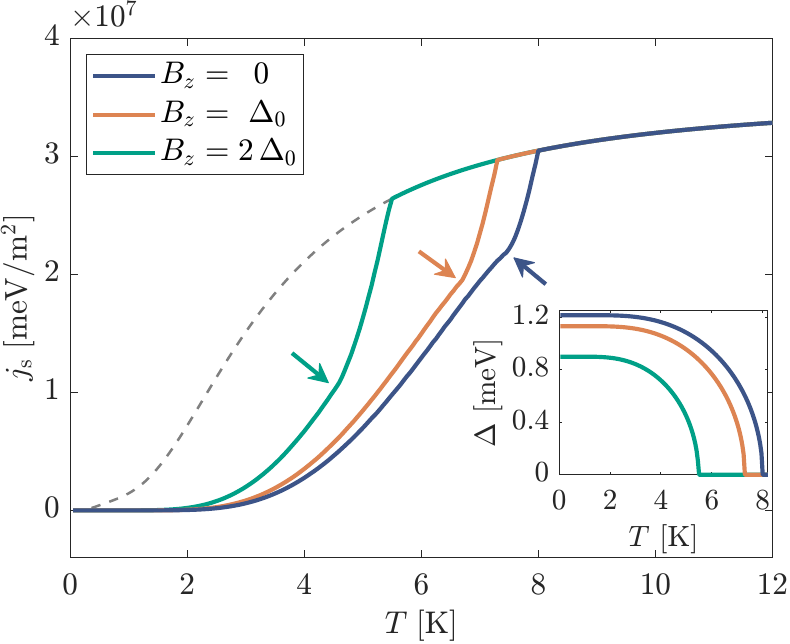}
	\caption{Temperature dependence of the injected spin current density $j_\text{s}$ in the Ising superconductor for different in-plane magnetic field strengths $B_z$.
		The colored arrows mark the kink below the superconducting transition on each curve.
		The gray dashed line represents the result for the normal state obtained by setting $\Delta=0$.
		The inset shows the temperature dependence of the superconducting gaps $\Delta$ for the corresponding in-plane magnetic field strengths.
		Here, the Ising SOC strength is fixed at $\beta_{\text{so}}=5\Delta_0$.}
	\label{Is_Bz}
\end{figure}

We define the injected spin current density as $j_\text{s} = I_\text{s}^\text{inj} / A_\text{sc}$.
Figure~\ref{Is_Bz} presents the temperature dependence of the injected spin current
density $j_\text{s}$ at different in-plane magnetic field strengths $B_z$.
The gray dashed line shows the normal-state reference calculated with $\Delta=0$.
At temperatures below approximately $2\,\text{K}$, the injected spin current density remains nearly zero.
In this regime, the population of thermally excited Bogoliubov quasiparticles is strongly suppressed by the superconducting gap.
As the temperature rises, the superconducting gap decreases and the population of thermally excited quasiparticles grows, resulting in an increase in the injected spin current density.
Furthermore, at a fixed temperature where the system remains superconducting under all three values of $B_z$, the injected spin current density increases with magnetic field strength.
At a fixed temperature, increasing the in-plane magnetic field strength reduces the superconducting gap [as shown by the gap curves in the inset of Fig.~\ref{Is_Bz}], thereby increasing the population of thermally excited quasiparticles and enhancing spin current injection.
For each $B_z$, the system enters the normal state at
the corresponding superconducting transition temperature
$T_\text{c}$ [see Table~\ref{table}].
For a chemical potential much larger than the relevant thermal, phonon, and spin-splitting energy scales, the imaginary part of the normal-state local advanced spin susceptibility is independent of the Ising SOC and $B_z$ [see Appendix~\ref{local_nm} for details]
\begin{equation}
	\text{Im} \chi^a_{\text{loc}}(\omega) 
	= 2\pi\hbar^2 \zeta^2 A_{\text{cell}}^2 \hbar \omega .
\end{equation}
Consequently, the spin current density curves at different $B_z$ coincide with the gray dashed normal-state reference curve.
Note that the $\bm{p}$-resolved spin susceptibility $\chi_{\bm{p}}^a(\omega)$ still depends on both the Ising SOC and $B_z$.

\begin{table}[b]
	\centering
	\caption{Superconducting transition temperatures, kink temperatures, and superconducting gaps at the kink points for different magnetic field strengths.}
	\label{table}
	\setlength{\tabcolsep}{16pt}
	\begin{tabular}{cccc}
		\toprule
		$B_z$  
		& $T_\text{c}\,[\text{K}]$ 
		& $T_{\text{k}}\,[\text{K}]$  
		& $\Delta(T_{\text{k}})\,[\text{meV}]$    \\
		\midrule
		$0$           & 8.00  & 7.45   & 0.538 \\
		$\Delta_0$    & 7.30  & 6.70   & 0.543 \\
		$2\Delta_0$   & 5.48  & 4.60   & 0.584 \\
		\bottomrule
	\end{tabular}
\end{table}

In addition to the pronounced kink at the superconducting transition temperature, a second kink appears as the temperature increases within the superconducting state, as indicated by the colored arrows in Fig.~\ref{Is_Bz}.
The kink temperatures $T_\text{k}$ and the corresponding superconducting gaps $\Delta(T_\text{k})$ for different $B_z$ are also listed in Table~\ref{table}.
To examine the origin of this kink, we compare the minimum quasiparticle excitation energy with the maximum phonon energy in our model.
The smaller positive eigenvalue of the BdG Hamiltonian is $E(\xi) = \sqrt{a(\xi) - \sqrt{b(\xi)}}$, where $a(\xi)$ and $b(\xi)$ are given in Eq.~\eqref{ab}.
The minimum value of $E(\xi)$ at $T_\text{k}$ is given by
\begin{equation}\label{Emin}
	E_{\text{min}}(T_\text{k},B_z) = \frac{\beta_{\text{so}} \Delta(T_\text{k})}
	{\sqrt{\beta_{\text{so}}^2 + B_z^2}} .
\end{equation}
Substituting the values of $B_z$ and the corresponding $\Delta(T_\text{k})$ listed in Table~\ref{table}, we find
\begin{equation}
	2E_{\text{min}} \simeq \hbar\omega_\text{max},
\end{equation}
where $\hbar\omega_\text{max} = 1.07 \,\text{meV}$ is the maximum phonon energy.
This correspondence is consistent with the onset of phonon-induced two-quasiparticle excitations as the excitation threshold $2E_{\text{min}}$ decreases into the phonon energy range with increasing temperature.
Such excitations can provide an additional contribution to the
injected spin current.

\begin{figure}[t]
	\centering
	\includegraphics[scale=0.61]{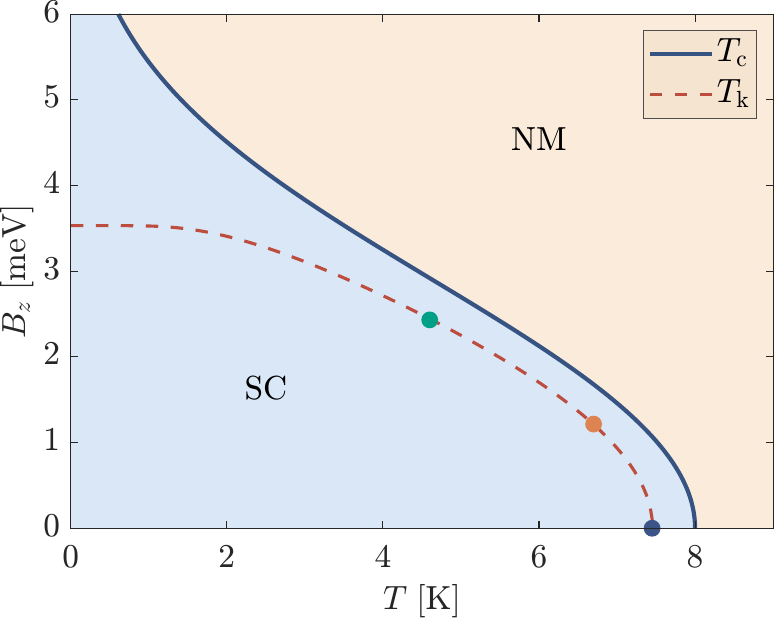}
	\caption{The $T$-$B_z$ phase diagram of the Ising superconductor.
	The dark blue solid line marks the phase boundary between the superconducting (SC) and normal metal (NM) states.
	The red dashed line marks the onset temperature $T_\text{k}$ of phonon-induced two-quasiparticle excitations as a function of $B_z$.
	The three colored points on the red dashed line correspond to the data in Table~\ref{table}.  }
	\label{Tk}
\end{figure}

To further elucidate the onset of phonon-induced excitations, we plot the $T$-$B_z$ phase diagram of the Ising superconductor and the corresponding onset temperature $T_\text{k}$ as a function of $B_z$, as shown in Fig.~\ref{Tk}.
The superconducting phase boundary is determined by solving the gap equation by considering $\Delta \rightarrow 0$.
Note that the critical in-plane magnetic field exhibits a logarithmic divergence as $T\to0$~\cite{Stefan2017}.
In an ideal Ising superconductor, the Ising SOC prevents $B_z$ from fully aligning the electron spins within the plane.
As the temperature decreases, a small but nonzero spin-singlet pairing component persists even under a strong in-plane magnetic field, leading to a divergence of the in-plane upper critical field in the zero-temperature limit.
Introducing intervalley scattering breaks this residual pairing and renders the upper critical field finite~\cite{Stefan2017, Tang2021}.
To obtain the $T_\text{k}$-$B_z$ curve, we first set $E_{\min}=\hbar\omega_{\max}/2$ in Eq.~\eqref{Emin} to determine the superconducting gap $\Delta$ at each value of $B_z$.
The corresponding temperature $T_\text{k}$ is then obtained by solving the superconducting gap equation~\cite{Tang2021, Hein2026}.
The region above (below) the red dashed line corresponds to the presence (absence) of phonon-induced two-quasiparticle excitations.
This result indicates that increasing $B_z$ suppresses the superconducting gap, thereby facilitating phonon-induced two-quasiparticle excitations and therefore reducing $T_\text{k}$.
The values of $T_\text{k}$ listed in Table~\ref{table} are also indicated by colored circles in Fig~\ref{Tk}.

\section{Summary and discussion}\label{summary}
In summary, we employ the nonequilibrium Green's function formalism to calculate the chiral-phonon-mediated spin current injection into an Ising superconductor.
We derive an expression for the injected spin current and show that it depends on the difference between the contributions from the two phonon branches with opposite circular polarizations.
We further investigate the temperature dependence of the injected spin current density at different in-plane magnetic field strengths.
In the superconducting state, the spin current density increases with temperature or in-plane magnetic field strength.
In addition to the kink at the superconducting transition, a second kink appears below the transition temperature.
The temperature of this second kink is consistent with the onset of phonon-induced two-quasiparticle excitations as the superconducting gap decreases with increasing temperature.
Such excitations can provide an additional contribution to the
injected spin current.
We also investigate the magnetic-field dependence of the kink temperature in a temperature--magnetic-field phase diagram.

\begin{acknowledgments}
G.\,L. is supported by the National Natural Science Foundation of China (Grant No.~12604081) and the Youth Innovation Team of Shaanxi Universities (Grant No.~25JP021).
\end{acknowledgments}

\appendix
\section{Wigner transformation of the equations of motion}\label{Wig_tran}
In this Appendix, we present the details of the Wigner transformation of the equations of motion in Eq.~\eqref{D_r<} term by term.
Because the transformation acts only on spatial coordinates, we suppress the frequency argument $\omega$ for simplicity.

\subsection{Frequency term $\hbar\omega D^{</r/a}_{ij,\lambda}$}
We express the phonon Green’s function in terms of $\bm{R}$ and $\bm{d}$, $D_{ij,\lambda} \rightarrow D_{\lambda}(\bm{R},\bm{d})$.
The Wigner transform of the frequency term then reads
\begin{equation}
	\mathcal{W} \big\{ \hbar\omega D^\ell_{\lambda}(\bm{R},\bm{d}) \big\}
	= \hbar\omega D^\ell_{\bm{q}\lambda}(\bm{R}).
\end{equation}
Here, $\ell$ labels the lesser, retarded, or advanced component, and $\mathcal{W}$ denotes the spatial Wigner transform.

\subsection{Free-phonon term $\sum_n \tilde{h}^\lambda_{in} D^\ell_{nj,\lambda}$}
For the term $\sum_n \tilde{h}^\lambda_{in} D^\ell_{nj,\lambda}$ with $\ell=>$, $<$, $r$, and $a$, we have
\begin{equation}\label{hD_1}
	\sum_n \tilde{h}^\lambda_{in} D^\ell_{nj,\lambda}
	= \frac{1}{2N_\text{ph}}\sum_{n,\bm{q}'}
	(\phi_i + \phi_n) 
	\hbar\omega_{\bm{q}'\lambda}
	e^{i\bm{q}'\cdot(\bm{r}_i-\bm{r}_n)}
	D^{\ell}_{nj,\lambda} .
\end{equation}
We introduce $\bm{d}_1 = \bm{r}_i - \bm{r}_n$ and $\bm{d}_2 = \bm{r}_n - \bm{r}_j$, so that $\bm{r}_i = \bm{R} + \bm{d}/2$ and $\bm{r}_n = \bm{R} + (\bm{d}_2 - \bm{d}_1)/2$.
Equation~\eqref{hD_1} then becomes
\begin{align}\label{hD_2}
	& \frac{1}{2N_\text{ph}}\sum_{\bm{q}',\bm{d}_1} 
	\Big[ 
	\phi \Big(\bm{R} + \frac{\bm{d}}{2} \Big) + 
	\phi\Big( \bm{R} + \frac{\bm{d}_2-\bm{d}_1}{2} \Big) \Big] 
	\nonumber\\
	& \hspace{23mm}  \times 
	\hbar\omega_{\bm{q}'\lambda}
	e^{i\bm{q}'\cdot \bm{d}_1}
	D^{\ell}_{\lambda} \Big( \bm{R}-\frac{\bm{d}_1}{2}, \bm{d}_2 \Big) .
\end{align}
We next expand the auxiliary field and phonon Green's function to first order in spatial gradients
\begin{subequations}
\begin{align}
	\phi \Big( \bm{R} + \frac{\bm{d}}{2} \Big) 
	& \simeq  \phi(\bm{R}) + 
	\frac{\bm{d}}{2} \cdot \nabla_{\bm{R}} \phi(\bm{R})  , \label{phi_1} 
	\\
	\phi\Big(\bm{R} + \frac{\bm{d}_2-\bm{d}_1}{2} \Big) & \simeq
	\phi(\bm{R}) + \frac{\bm{d}_2-\bm{d}_1}{2} \cdot \nabla_{\bm{R}} \phi(\bm{R}),   \label{phi_2} 
	\\
	D^{\ell}_{\lambda} \Big( \bm{R}-\frac{\bm{d}_1}{2}, \bm{d}_2 \Big) & \simeq D^{\ell}_{\lambda}(\bm{R}, \bm{d}_2)
	- \frac{\bm{d}_1}{2} \cdot \nabla_{\bm{R}} D^{\ell}_{\lambda}(\bm{R}, \bm{d}_2) . \label{D} 
\end{align}	
\end{subequations}
Substituting Eqs.~\eqref{phi_1}-\eqref{D} into Eq.~\eqref{hD_2} yields
\begin{align}\label{hD}
	& \frac{1}{2N_\text{ph}}\sum_{\bm{q}',\bm{d}_1} \hbar\omega_{\bm{q}'\lambda}
	e^{i\bm{q}'\cdot \bm{d}_1} 
	\nonumber\\
	& \quad\times
	\Big[
	2\phi(\bm{R})  D^{\ell}_{\lambda}(\bm{R}, \bm{d}_2)
	- \phi(\bm{R}) \bm{d}_1 \cdot \nabla_{\bm{R}} 
	D^{\ell}_{\lambda}(\bm{R}, \bm{d}_2) 
	\nonumber\\ & \qquad +  
	\bm{d}_2 \cdot \nabla_{\bm{R}} \phi(\bm{R}) D^{\ell}_{\lambda}(\bm{R}, \bm{d}_2)   \Big]  .
\end{align}
We now transform the three terms in Eq.~\eqref{hD} separately
\begin{subequations}
\begin{align}
	w_1 & = 
	\frac{\phi(\bm{R})}{N_\text{ph}} \sum_{\bm{q}',\bm{d}_1} 
	\hbar\omega_{\bm{q}'\lambda} 
	e^{i(\bm{q}'-\bm{q})\cdot\bm{d}_1} 
	\sum_{\bm{d}_2} e^{-i\bm{q}\cdot\bm{d}_2} 
	D^{\ell}_{\lambda}(\bm{R}, \bm{d}_2)
	\nonumber\\
	& \quad = \phi(\bm{R}) \hbar\omega_{\bm{q}\lambda} 
	D^{\ell}_{\bm{q}\lambda}(\bm{R}), \label{w_p1}
	\\
	w_2 & = -\frac{\phi(\bm{R})}{2N_\text{ph}} 
	\sum_{\bm{q}'} \hbar\omega_{\bm{q}'\lambda} 
	\nonumber\\*
	& \quad \times
	\sum_{\bm{d}_1} \bm{d}_1 e^{i(\bm{q}'-\bm{q})\cdot\bm{d}_1}
	\cdot \nabla_{\bm{R}} \sum_{\bm{d}_2} e^{-i\bm{q}\cdot\bm{d}_2} 
	D^{\ell}_{\lambda}(\bm{R}, \bm{d}_2) 
	\nonumber\\*
	& = -\frac{i\hbar}{2} \phi(\bm{R}) 
	\nabla_{\bm{q}} \omega_{\bm{q}\lambda} \cdot
	\nabla_{\bm{R}} D^{\ell}_{\bm{q}\lambda}(\bm{R}) , \label{w_p2}
	\\
	w_3 & = \frac{\nabla_{\bm{R}} \phi(\bm{R})}{2N_\text{ph}} \cdot
	\sum_{\bm{q}'} \hbar\omega_{\bm{q}'\lambda}
	\nonumber\\ & \quad \times
	\sum_{\bm{d}_1} e^{i(\bm{q}'-\bm{q})\cdot\bm{d}_1}
	\sum_{\bm{d}_2} e^{-i\bm{q}\cdot\bm{d}_2} 
	\bm{d}_2 D^{\ell}_{\lambda}(\bm{R}, \bm{d}_2) \nonumber\\
	& = \frac{i\hbar}{2} \omega_{\bm{q}\lambda} \nabla_{\bm{R}} \phi(\bm{R}) \cdot
	\nabla_{\bm{q}} D^{\ell}_{\bm{q}\lambda}(\bm{R}) . \label{w_p3}
\end{align}
\end{subequations}
Adding Eqs.~\eqref{w_p1}-\eqref{w_p3} gives
\begin{align}
	& \mathcal{W} 
	\Big\{ \sum_n \tilde{h}^\lambda_{in} D^\ell_{nj,\lambda} \Big\}
	= w_1 + w_2 + w_3
	\nonumber\\
	& \quad = \phi(\bm{R}) \hbar\omega_{\bm{q}\lambda} 
	D^{\ell}_{\bm{q}\lambda}(\bm{R})
	-\frac{i\hbar}{2} \phi(\bm{R}) 
	\nabla_{\bm{q}} \omega_{\bm{q}\lambda} \cdot
	\nabla_{\bm{R}} D^{\ell}_{\bm{q}\lambda}(\bm{R}) 
	\nonumber\\
	& \qquad + \frac{i\hbar}{2} \omega_{\bm{q}\lambda} \nabla_{\bm{R}} \phi(\bm{R}) \cdot
	\nabla_{\bm{q}} D^{\ell}_{\bm{q}\lambda}(\bm{R}) .
\end{align}

\subsection{Kronecker-delta term $\delta_{ij}$}
The Wigner transform of the Kronecker delta is
\begin{equation}
	\mathcal{W}\{ \delta_{ij} \} \rightarrow 
	\mathcal{W}\{ \delta_{\bm{d},0} \} 
	= \sum_{\bm{d}} e^{-i\bm{q}\cdot\bm{d}} \delta_{\bm{d},0} = 1.
\end{equation}

\subsection{Self-energy terms $\sum_n \Sigma^{\ell_1}_{in,\lambda} D^{\ell_2}_{nj,\lambda}$}
For the term $\sum_n \Sigma^{\ell_1}_{in,\lambda} D^{\ell_2}_{nj,\lambda}$ with $\ell_{1,2} = >$, $<$, $r$, and $a$, expressing it in terms of the center and relative coordinates yields
\begin{equation}\label{T4}
	\sum_{\bm{d}_1}  \Sigma^{\ell_1}_{\lambda} 
	\Big( \bm{R} + \frac{\bm{d}_2}{2}, \bm{d}_1 \Big)
	D^{\ell_2}_{\lambda}
	\Big( \bm{R} - \frac{\bm{d}_1}{2}, \bm{d}_2 \Big) .
\end{equation}
Similarly, we expand the self-energy and Green's function to first order in spatial gradients
\begin{subequations}
\begin{align}
	\Sigma^{\ell_1}_{\lambda} \Big(\bm{R} + \frac{\bm{d}_2}{2}, \bm{d}_1\Big) & \simeq 
	\Sigma^{\ell_1}_{\lambda} (\bm{R}, \bm{d}_1) 
	+ \frac{\bm{d}_2}{2} \cdot
	\nabla_{\bm{R}} \Sigma^{\ell_1}_{\lambda} (\bm{R}, \bm{d}_1), 
	\\
	D^{\ell_2}_{\lambda} \Big(\bm{R} - \frac{\bm{d}_1}{2}, \bm{d}_2\Big) & \simeq
	D^{\ell_2}_{\lambda}(\bm{R}, \bm{d}_2) 
	- \frac{\bm{d}_1}{2} \cdot
	\nabla_{\bm{R}} D^{\ell_2}_{\lambda}(\bm{R}, \bm{d}_2) .
\end{align}	
\end{subequations}
To first order in spatial gradients, the Wigner transform of Eq.~\eqref{T4} is given by
\begin{align}\label{SD_1}
	& \sum_{\bm{d}_1 \bm{d}_2} e^{-i\bm{q}\cdot(\bm{d}_1 + \bm{d}_2)}
	\Big\{
	\Sigma^{\ell_1}_{\lambda}(\bm{R}, \bm{d}_1) 
	D^{\ell_2}_{\lambda}(\bm{R}, \bm{d}_2)
	\nonumber\\
	& \qquad - \frac{1}{2} \Sigma^{\ell_1}_{\lambda}(\bm{R}, \bm{d}_1) 
	\big[ \bm{d}_1 \cdot \nabla_{\bm{R}} 
	D^{\ell_2}_{\lambda}(\bm{R}, \bm{d}_2) \big]
	\nonumber\\ 
	& \qquad + \frac{1}{2} \big[  \bm{d}_2 \cdot \nabla_{\bm{R}} 
	\Sigma^{\ell_1}_{\lambda}(\bm{R}, \bm{d}_1)  \big]
	D^{\ell_2}_{\lambda}(\bm{R}, \bm{d}_2) \Big\}  .
\end{align}
The three terms in Eq.~\eqref{SD_1} give
\begin{subequations}
\begin{align}
	w_1 & = \sum_{\bm{d}_1 \bm{d}_2} 
	e^{-i\bm{q}\cdot(\bm{d}_1 + \bm{d}_2)}
	\Sigma^{\ell_1}_{\lambda}(\bm{R}, \bm{d}_1) 
	D^{\ell_2}_{\lambda}(\bm{R}, \bm{d}_2)
	\nonumber\\
	& = \Sigma^{\ell_1}_{\bm{q}\lambda}(\bm{R})
	D^{\ell_2}_{\bm{q}\lambda}(\bm{R}) ,  
	\\
	w_2 & = - \frac{1}{2} \sum_{\bm{d}_1 \bm{d}_2} 
	e^{-i\bm{q}\cdot(\bm{d}_1 + \bm{d}_2)}
	\Sigma^{\ell_1}_{\lambda}(\bm{R}, \bm{d}_1) 
	\big[ \bm{d}_1 \cdot \nabla_{\bm{R}} 
	D^{\ell_2}_{\lambda}(\bm{R}, \bm{d}_2) \big]
	\nonumber\\
	& = -\frac{1}{2} \sum_{\alpha} 
	\sum_{\bm{d}_1} e^{-i\bm{q}\cdot\bm{d}_1}
	\Sigma^{\ell_1}_{\lambda}(\bm{R}, \bm{d}_1) d_{1,\alpha} 
	\nonumber\\
	& \hspace{25mm} \times
	\partial_{R_{\alpha}} \sum_{\bm{d}_2} e^{-i\bm{q}\cdot\bm{d}_2}
	D^{\ell_2}_{\lambda}(\bm{R}, \bm{d}_2) 
	\nonumber\\
	& = -\frac{i}{2} \sum_{\alpha} \frac{\partial}{\partial q_{\alpha}} 
	\Big[\sum_{\bm{d}_1} e^{-i\bm{q}\cdot\bm{d}_1} 
	\Sigma^{\ell_1}_{\lambda}(\bm{R}, \bm{d}_1) \Big]
	\nonumber\\
	& \hspace{25mm} \times
	\partial_{R_{\alpha}} \sum_{\bm{d}_2} e^{-i\bm{q}\cdot\bm{d}_2}
	D_{\lambda}^{\ell_2}(\bm{R}, \bm{d}_2) 
	\nonumber\\
	& = -\frac{i}{2} 
	\nabla_{\bm{q}} \Sigma^{\ell_1}_{\bm{q}\lambda}(\bm{R})  \cdot
	\nabla_{\bm{R}} D^{\ell_2}_{\bm{q}\lambda}(\bm{R}) ,
	\\
	w_3 & = \frac{1}{2} \sum_{\bm{d}_1 \bm{d}_2} 
	e^{-i\bm{q}\cdot(\bm{d}_1 + \bm{d}_2)}
	\big[  \bm{d}_2 \cdot \nabla_{\bm{R}} 
	\Sigma^{\ell_1}_{\lambda}(\bm{R}, \bm{d}_1)  \big]
	D^{\ell_2}_{\lambda}(\bm{R}, \bm{d}_2) 
	\nonumber\\
	& = \frac{i}{2} 
	\nabla_{\bm{R}} \Sigma^{\ell_1}_{\bm{q}\lambda}(\bm{R})  
	\cdot \nabla_{\bm{q}} 
	D^{\ell_2}_{\bm{q}\lambda}(\bm{R}).
\end{align}	
\end{subequations}
Thus the Wigner transform of this term is
\begin{align}
	& \mathcal{W}
	\Big\{ \sum_n \Sigma^{\ell_1}_{in,\lambda} D^{\ell_2}_{nj,\lambda} \Big\} = w_1 + w_2 + w_3 \nonumber\\
	& = \Sigma^{\ell_1}_{\bm{q}\lambda}(\bm{R}) 
	D^{\ell_2}_{\bm{q}\lambda}(\bm{R})
	+ \frac{i}{2} \big[ 
	\nabla_{\bm{R}} \Sigma^{\ell_1}_{\bm{q}\lambda}(\bm{R})  \cdot
	\nabla_{\bm{q}} D^{\ell_2}_{\bm{q}\lambda}(\bm{R})
	\nonumber\\
	& \quad - 
	\nabla_{\bm{q}} \Sigma^{\ell_1}_{\bm{q}\lambda}(\bm{R})  \cdot
	\nabla_{\bm{R}} D^{\ell_2}_{\bm{q}\lambda}(\bm{R})
	\big] .
\end{align}
Combining the above Wigner-transformed results and introducing the following star-product notation
\begin{equation}
	A \star B = A B + \frac{i}{2} 
	\big( \nabla_{\bm{R}} A \cdot \nabla_{\bm{q}} B -
	\nabla_{\bm{q}} A \cdot \nabla_{\bm{R}} B \big) ,
\end{equation}
we can obtain the following Dyson and Keldysh equations
\begin{subequations}
\begin{gather}
	\big[ \hbar\omega  - \phi \hbar\omega_{\bm{q}\lambda} -
	\Sigma^{r/a}_{\bm{q}\lambda}(\bm{R}) \big] 
	\star D^{r/a}_{\bm{q}\lambda}(\bm{R}) 
	= 1 , 
	\\
	D^{<}_{\bm{q}\lambda}(\bm{R}) = 
	D^r_{\bm{q}\lambda}(\bm{R}) \star
	\Sigma^{<}_{\bm{q}\lambda}(\bm{R}) \star 
	D^{a}_{\bm{q}\lambda}(\bm{R}) .  
\end{gather}
\end{subequations}

\section{Local susceptibility of the normal state}\label{local_nm}
We next derive the local advanced susceptibility in the normal state.
Since the anomalous Green's functions vanish, Eq.~\eqref{chi_a_1} reduces to
\begin{align}\label{sus_nm1}
	\chi_{\bm{p}}^a(\omega) 
	& = -\frac{i\hbar^2}{N_{\text{S}}} \sum_{\bm{k},\nu} \int_{-\infty}^{\infty} 
	\frac{dE}{2\pi} 
	\big[
	G_{\bm{k}+\bm{p}\nu,\downarrow\downarrow}^a(\hbar\omega+E) G_{\bm{k}\nu,\uparrow\uparrow}^<(E) 
	\nonumber\\ 
	& \quad + 
	G_{\bm{k}+\bm{p}\nu,\downarrow\downarrow}^<(\hbar\omega+E) G_{\bm{k}\nu,\uparrow\uparrow}^r(E)   \big] .
\end{align}
Here the retarded/advanced Green's functions are defined as
\begin{equation}
	\mathbf{G}^{r/a}_{\bm{k}\nu}(E) = 
	\big[ (E\pm i\eta) {\mathbf{I}}_2 - h_0(\bm{k},\nu) \big]^{-1} .
\end{equation}
Following a derivation analogous to that for the superconducting state, we obtain
\begin{align}
	\chi_{\bm{p}}^a(\omega) 
	& = \frac{2\hbar^2}{N_{\text{S}}} \sum_{\bm{k}} \sum_{n,m} 
	\nonumber\\
	& \quad \times 
	r^{(11)}_{\bm{k},n}  r^{(22)}_{\bm{k}+\bm{p},m}
	\frac{f(E_{\bm{k},n}) - f(E_{\bm{k}+\bm{p},m})}
	{\hbar\omega + E_{\bm{k},n} - E_{\bm{k}+\bm{p},m} - i\eta} ,
\end{align}
where $E_{\bm{k},n} = \xi_{\bm{k}} + n \sqrt{ \beta_{\text{so}}^2 + B_z^2 }$ and 
\begin{subequations}
\begin{align}
	r^{(11)}_{\bm{k},n} & = \frac{1}{2} 
	\bigg( 1 - n\frac{B_z}{ \sqrt{ \beta_{\text{so}}^2 + B_z^2 }} \bigg) , \\
	r^{(22)}_{\bm{k}+\bm{p},m} & = \frac{1}{2} 
	\bigg( 1 + m\frac{B_z}{ \sqrt{ \beta_{\text{so}}^2 + B_z^2 }} \bigg) 
\end{align}	
\end{subequations}
with $n,m = \pm 1$.
Using the definition of the local spin susceptibility given in Eq.~\eqref{loc_def} and converting the wavevector summations into energy integrals, we obtain
\begin{align}
	&  \text{Im} \chi^a_{\text{loc}}(\omega) 
	= 2\pi\hbar^2 \zeta^2 A_{\text{cell}}^2  \sum_{n,m}
	r^{(11)}_{\xi,n}  r^{(22)}_{\xi',m}
	\int_{-\mu}^{\infty} d\xi \int_{-\mu}^{\infty} d\xi'
	\nonumber\\   
	&  \quad \times
	\big[ f(E_{\xi,n}) - f(E_{\xi',m}) \big]
	\delta(\hbar\omega + E_{\xi,n} - E_{\xi',m}) .
\end{align}
Here the $\delta$ function imposes the constraint
\begin{equation}
	\xi' = \hbar\omega +  \xi + (n-m)\sqrt{ \beta_{\text{so}}^2 + B_z^2 } .
\end{equation}
Letting $\tilde{E} = \xi + n \sqrt{ \beta_{\text{so}}^2 + B_z^2 } $ and considering
$\mu \gg \max \big\{ k_\text{B} T, |\hbar\omega|, \sqrt{\beta_{\text{so}}^2 + B_z^2} \big\}$,
we have
\begin{align}
	\text{Im} \chi^a_{\text{loc}}(\omega) 
	& = 2\pi\hbar^2 \zeta^2 A_{\text{cell}}^2  \sum_{n,m}
	r^{(11)}_{\xi,n}  r^{(22)}_{\xi',m}
	\nonumber\\   
	&  \quad \times
	\int_{-\infty}^{\infty} d \tilde{E}
	\big[ f(\tilde{E}) - f(\tilde{E} + \hbar\omega) \big] .
\end{align}
Using the identities
\begin{subequations}
\begin{gather}
	\sum_{n,m} r^{(11)}_{\xi,n}  r^{(22)}_{\xi',m} 
	= \Big(\sum_n  r^{(11)}_{\xi,n}\Big)  \Big(\sum_m r^{(22)}_{\xi',m}\Big) = 1 , \\
	\int_{-\infty}^{\infty} d \tilde{E}
	\big[ f(\tilde{E}) - f(\tilde{E} + \hbar\omega) \big] = \hbar\omega ,
\end{gather}	
\end{subequations}
we finally obtain
\begin{equation}
	\text{Im} \chi^a_{\text{loc}}(\omega) 
	= 2\pi\hbar^2 \zeta^2 A_{\text{cell}}^2 \hbar \omega .
\end{equation}
Thus, the imaginary part of the normal-state local advanced spin susceptibility is independent of both the Ising SOC and $B_z$.

\bibliography{reference.bib}

\end{document}